# Helical-Phase Distinguishability for the Phase Distribution of Laguerre-Gaussian Beams

Arturo Pazmino[1,a)], Peter Iza[1,2,b)], Manuel S. Alvarez-Alvarado[3,c)], Erick Lamilla[1,d)]

[1]Escuela Superior Politécnica del Litoral, ESPOL, Departamento de Física, Camus Gustavo Galindo km 30.5 Vía Perimetral, P.O. Box 09-01-5863, Guayaquil, Ecuador
[2]Center of Research and Development in Nanotechnology, CIDNA, Escuela Superior Politécnica del Litoral, ESPOL, Departamento de Física, Camus Gustavo Galindo km 30.5 Vía Perimetral, P.O. Box 09-01-5863, Guayaquil, Ecuador
[3]Escuela Superior Politécnica del Litoral, ESPOL, Facultad de Ingeniería Eléctrica y Computación, FIEC, Camus Gustavo Galindo km 30.5 Vía Perimetral, P.O. Box 09-01-5863, Guayaquil, Ecuador

[a)] Corresponding author: apazmino@espol.edu.ec
[b)] Electronic email:piza@espol.edu.ec
[c)] Electronic email: mansalva@espol.edu.ec
[d)] Electronic email: ealamill@espol.edu.ec

**Abstract.** In the past few years, orbital angular momentum beams have contributed significantly to many applications in the optical-related fields due to their unique phase characteristics. Here, a novel approach about phase-helices structure for Laguerre-Gaussian beams carrying orbital angular momentum is presented. This proposal is based on numerical simulations to track the rotation of each helix in the phase and shows an impact on the rotational speed of the Laguerre-Gaussian phase distribution by using the characteristic of distinguishability of each helix. If the helices in the phase of a vortex cannot be distinguished, then the angular rotation speed of the phase distribution can be seen as independent of the azimuthal index l, due to its rotational symmetry, and it is called $\omega_{indistinguishable}$. Oppositely, when the helices are distinguished from each other, the rotational speed is affected by this azimuthal index, and it becomes $\omega_{distinguishable}$. This distinguishable theory can build a pathway to applications in the field of optical communications (as a coding system) and particle manipulation (changing the dynamics of off-beam trapped particles).

## INTRODUCTION

Electromagnetic waves can be represented as the synchronized oscillation of electric and magnetic field propagating at the speed of light in vacuum, these waves have two distinct physical properties: spin angular momentum (SAM) and orbital angular momentum (OAM) [1]. According to Beth [2], an optical beam propagates with a spin angular momentum due to its polarization. However, in 1992 Alen *et al.* explained, for the first time, the propagation of the optical beam with an entire OAM related to the structure of the wavefront and spatial distribution of the electromagnetic radiation [3]. Since then, the study of the OAM in light has become more attractive due to its important contributions on applications in the fields of particle manipulation using light through optical tweezers [4–7], codification of information [8–10], improvement of borders in the image processing [11], super-resolution imaging [12], laser processing [13], and applications related to spatial multiplexing in optical communications [14–19].

Studies based on group velocity changes in Laguerre-Gaussian (LG) beams with OAM were conceptualized by Allen, resulting in the theoretical work presented in 1994, where the azimuthal Doppler shift in LG beams with OAM is proposed. They show that, an atom moving in a LG vortex beam exhibits an azimuthal shift in the resonant frequency, adding an azimuthal component of velocity. This prediction played a significant role in particle manipulation and its impact in the group and phase velocity of vortex beams [20].

Literature presents expanded studies of optical beams carrying OAM, for instance, Luo *et al.* theoretically demonstrated that rotational Doppler effect in some materials is unreversed, due to the combined contributions of negative phase velocity and inverse screw of wavefront [21]. Lavery *et al.* reported an experimental observation of spinning objects using the orbital angular momentum of light scattering, where the degree of OAM enhancement of rotation direction is a function of the experimental conditions [22]. The concept of angular rotation with light was also explored in the form of angular acceleration for Schulze *et al.*, where a class of the light field with angular acceleration during propagation is studied. In that work, angular accelerating light fields and conservation of angular momentum through an energy exchange mechanism across the optical field is discussed [23]. Additionally, there has been works about petal-like intensity structure from the superposition of LG beams due to its implications with quantum information [24,25]. This petal-like structure has been used to experimentally measure, for the first time, tiny velocities [26].

Regarding monochromatic plane waves, the propagation velocity, referred to as the phase velocity in vacuum, is a constant; but any deviation from the plane wave constraint can lead to a propagation velocity different from *c* (speed of light in vacuum). Bouchard *et al.* show that twisted light pulses exhibit subluminal velocities in vacuum, being 0.1% slower relative to *c* [27]. This research invites us to draw a difference between the group velocity and the phase velocity for light beams with a twisted wavefront. Generally, for a plane wave propagating along with a nondispersive medium, both group and phase velocity take the same value (i.e., $v_g = v_{ph} = c/n$), being *n* the refractive index of the medium. Nevertheless, concerning twisted light (also known as vortex beam), the phase and group velocities can differ in magnitude. Under specific conditions, group velocity in optical beams can be superluminal [28,29], and even have a negative velocity propagation [30,31].

The light with OAM is associated with the spatial structure of the optical field. This structure is helical in both, wavefront and phase distribution. For the phase distribution case, an optical beam with a helical phase, that depends on *exp(-ilϕ)*, carries an orbital angular momentum independent of the polarization state. The angle $\phi$ is the azimuthal coordinate in the beam's cross section, and *l*, the topological charge (or azimuthal index) of the optical beam, takes integer and fractional values [32,35]. The most common form of a helical phased beam is the Laguerre-Gaussian (LG) [3]. Helical wavefronts are also observed in Bessel beams [36], Mathieu beams [37], and Ince–Gaussian beams [38].

Due to the impact of the OAM regarding the helical structure of vortex beam's phase distribution in the past few years, this work proposed an innovative philosophy to study the well-known rotational velocity based on a "distinguishable helical-phase" theory, not mentioned elsewhere in the literature, to the author's knowledge. This approach may build a pathway for new mathematical developments for future applications in optical communications.

## LAGUERRE-GAUSSIAN VORTEX BEAM OVERVIEW

In free space, an optical field can be described by a cylindrical wave which is characterized by Laguerre polynomials. An optical beam propagating in the z-direction in free space can be expressed as:

$$E(r,\phi,z) = U(r,\phi,z) \exp(-ikz) \tag{1}$$

where $U(r,\phi,z)$ is the amplitude of beam in cylindrical coordinates. The scalar optical field satisfies the Helmholtz equation, and the paraxial wave equation is derived when the second derivative in the z-direction is ignored. Then, the cylindrical symmetric solutions Laguerre–Gaussian beams are given by [3]:

$$U(r,\phi,z) = \sqrt{\frac{2p!}{\pi(p+|l|)!}} \frac{1}{\omega(z)} \left(\frac{r\sqrt{2}}{\omega(z)}\right)^{|l|} L_p^{|l|}\left(\frac{2r^2}{\omega^2(z)}\right) exp\left(-\frac{r^2}{\omega^2(z)}\right) exp\left(-i\left(l\phi + \frac{kr^2}{2R_z} - (2p+|l|+1)tan^{-1}\left(\frac{z}{z_0}\right)\right)\right) \tag{2}$$

where $z_0 = \pi\omega_0^2/\lambda$ is the Rayleigh range, $R_z = z(1 + (z/z_0)^2)$ is the radius of curvature, $\lambda$ is the wavelength and $k = 2\pi/\lambda$ is the wave number, $\omega_z = \omega_0\sqrt{1 + (z/z_0)^2}$ is the beam radius with $\omega_0$ being the beam waist, $L_p^{|l|}(x)$ is the associated Laguerre polynomial.

In the literature, the term $tan^{-1}(z/z_0)$ is known as the Gouy's phase and is multiplied with the mode order $(2p + |l| + 1)$. All LG beams are characterized by two indexes: the radial index $p$ and the azimuthal index $l$, also called topological charge for vortex beams, denoting the orbital angular momentum (in units of $\hbar$). The phase distribution of a LG vortex beam is shown in the following equation:

$$\Phi(r,\phi,z) = kz + l\phi + \frac{kr^2}{2R_z} - (2p + |l| + 1)tan^{-1}\left(\frac{z}{z_0}\right) \qquad (3)$$

then, the group velocity (rate at which the envelope of the beam's amplitude propagates through space) and phase velocity (rate at which a point of constant phase propagates through space) is given by [39,40]:

$$v_g = \frac{1}{|\nabla[\delta_\omega \Phi(r,\omega)]|} \qquad (4)$$

and

$$v_{ph} = \frac{\omega}{|\nabla \Phi(r)|} \qquad (5)$$

where $\omega$ is the angular frequency.

The Gouy's phase shift term, in the longitudinal phase of a LG beam, is an intrinsic characteristic of an electromagnetic field that is spatially confined in the transverse plane of the propagation. An important observation is that the phase velocity can be slightly bigger than the speed of light in free space (superluminal velocity). It is relevant to mention that such fact does not violate special relativity, because any signal information sent by this LG vortex beam (or any other optical beam) travels at the speed of light (or even at lower speed) [41]. For a LG vortex beam, information will be transmitted at its group velocity $v_g$, that is slightly smaller than the speed of light (subluminal velocity) [40].

## SPEED ROTATION OF LAGUERRE-GAUSSIAN PHASE DISTRIBUTION

Moving one step further by analyzing the rotation velocity of the phase distribution of a LG beam through simulating the argument of the full LG Eq. (1) and analyzing its rotation at different values of the propagation axis. Fig. 1 shows the track rotation of the beam phase distribution for different values of topological charge $l$ and radial index $p$ ($LG_p^l$), with a beam waist $\omega_0 = 5.0$ mm and wavelength $\lambda = 633$ nm. Fig. 1a, is the rotation of a LG beam phase for $LG_0^1$, with a phase distribution from $-\pi$ to $\pi$ (Fig. 1a.i), its rotation is counterclockwise due to the term $\exp(-ikz)$. The phase distribution starts at $z = 0$ (Fig. 1a.ii), it has a half rotation at $z = \lambda/2$ (Fig. 1a.iii) and a full rotation at $z = \lambda$ (Fig. 1a.iv). Then, the period of this phase rotation is $T = \lambda/v_0$, where $v_0$ is a linear velocity at which the phase moves along the axis propagation.

Fig. 1b, shows the rotation phase of a $LG_0^2$ beam, the phase distribution has two helices, blue-green color map to the left and orange color map to the right, both with a phase distribution from $-\pi$ to $\pi$, in Fig. 1b.i. The helical-phase distribution starts at $z = 0$ (Fig. 1b.ii), has a half rotation at $z = \lambda$ (Fig. 1b.iii), and a full rotation at $z = 2\lambda$ (Fig. 1b.iv). Fig. 1c and 1d, shows the rotation phase for a $LG_1^1$ and $LG_2^2$ beams, which follow the same behavior as Fig. 1a and 1b respectively.

In general, from Fig. 1, the radial index $p$ does not affect the phase angular rotation speed, and the period of the rotation only depends on the topological charge $l$. Moreover, analyzing step by step the period of the helical-phase (hp) rotation of a LG beams from this figure, for $l = 1$ the period is $T_{hp} = \lambda/v_0$ (Fig. 1a and 1c), for $l = 2$ the period is $T_{hp} = 2\lambda/v_0$ (Fig. 1b and 1d); going further, for $l = 3$ the period is $T_{hp} = 3\lambda/v_0$; and so on, therefore, generalizing the expression, the helical-phase rotation period would be $T_{hp} = l\lambda/v_0$. Then, the rotation speed of the helical-phase distribution of a LG vortex beam is given by

$$\omega_{hp} = \frac{2\pi v_0}{\lambda l} = \frac{k v_0}{l} \qquad (6)$$

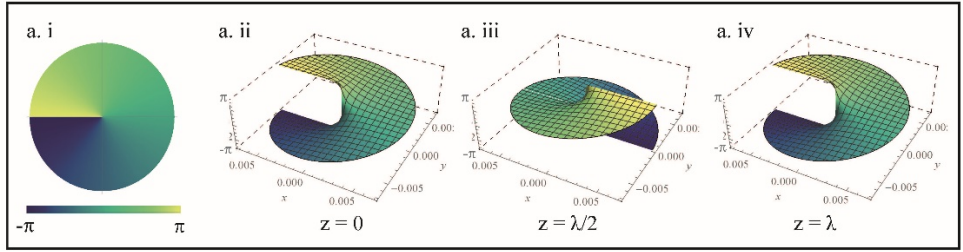
(a) Tracking of angular rotation for a LG01 phase distribution

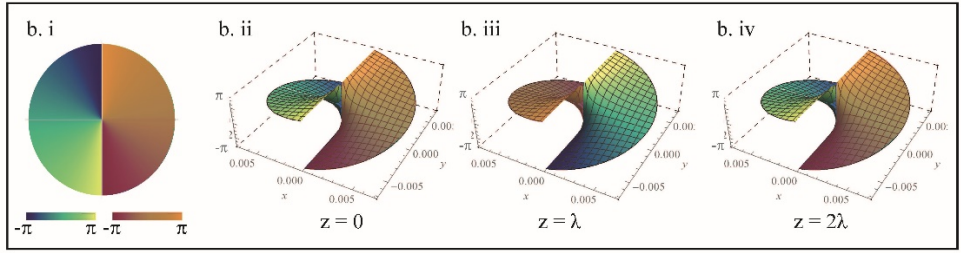
(b) Tracking of angular rotation for a LG02 phase distribution

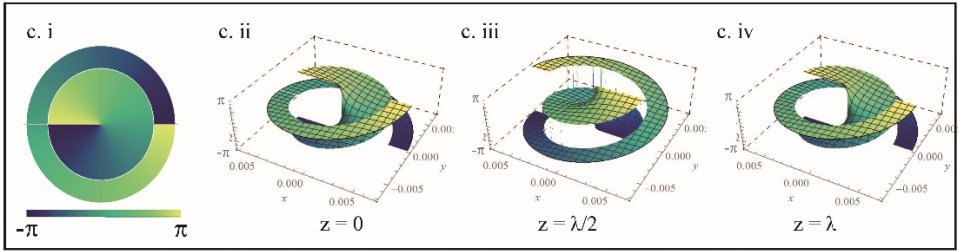
(c) Tracking of angular rotation for a LG11 phase distribution

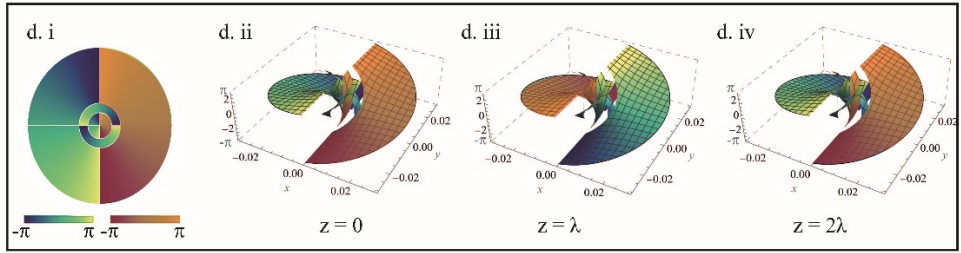
(d) Tracking of angular rotation for a LG22 phase distribution

**FIGURE 1.** Tracking of the rotation of Laguerre-Gaussian phase distribution ($LG_p^l$) for a) $LG_0^1$, b) $LG_0^2$, c) $LG_1^1$, and d) $LG_2^2$; using a fixed beam waist $\omega_0 = 5.0$ mm and wavelength $\lambda = 633$ nm, at different values of the axis propagation, $z$. For $l = 2$, there are two helices colored in different gradient colors (blue-green and orange) for rotational tracking purposes.

## DISTINGUISHABILITY FEATURE OF A HELICAL-PHASE DISTRIBUTION

A known feature, observed in Fig. 1, is that the number of helices is only given by the topological charge $l$, and the characteristic of distinguishability of the helical-phase distribution in the LG vortex beam can be inferred from it. This distinguishability characteristic is better illustrated in Fig. 2, in which the phase distribution is plotted for a LG vortex beam of waist $\omega_0 = 5.0$ mm and wavelength $\lambda = 633$ nm, for $p = 0$ and $l = 3$, at different values of the axis propagation.

In Fig. 2a, the helical-phase distribution is plotted using the same color, such that helices are indistinguishable from one to another. Then, at $z = 0$ the phase has certain distribution (Fig. 2a.i) that is repeated at $z = \lambda$ (Fig. 2a.ii), $2\lambda$ (Fig. 2a.iii) and $3\lambda$ (Fig. 2a.iv), therefore, the apparent period would be $T = \lambda/v_0$, giving an apparent rotation speed of $\omega_{indistinguishable} = kv_0$, due to its rotational symmetry.

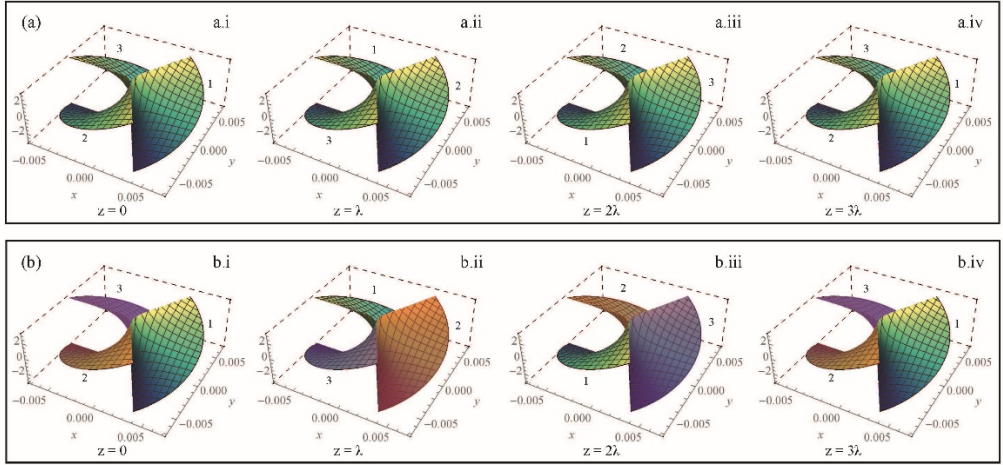

**FIGURE 2.** Tracking of the rotation of the LG phase distribution for $p = 0$ and $l = 3$ using a beam waist $\omega_0 = 5.0$ mm and wavelength $\lambda = 633$ nm, at different values of the axis propagation, $z = 0, \lambda, 2\lambda,$ and $3\lambda$. a) All helices in the phase distribution have the same color, making them indistinguishable from one another. b) Each helix in the phase structure has a different color, making them distinguishable from one to another.

On the other hand, if each helix in the phase distribution is plotted using different colors, as it is shown in Fig. 2b, the real rotation of the phase distribution can be tracked. Now the helical-phase structure is distinguishable from one to another, the phase distribution starts at $z = 0$ (Fig. 2b.i), for $z = \lambda$ the entire phase has rotated 120 degrees (Fig. 2b.ii), and for $z = 2\lambda$ it has rotated 240 degrees (Fig. 2b.iii). At $z = 3\lambda$ the phase rotates a full cycle (Fig. 2b.iv); therefore, the rotation period would be $T = l\lambda/v_0$, giving a rotational velocity of $\omega_{distinguishable} = kv_0/l$ as in Eq. (6).

To emphasize the latest finding, the rotational velocities $\omega_{indistinguishable}$ and $\omega_{distinguishable}$ are plotted in Fig. 3, using 633 nm of wavelength, beam waist of $\omega_0 = 5.0$ mm, $l = 1,2,3,4$, and $p = 0$. When the helical-phase is indistinguishable (dashed lines), the rotational velocity of the phase distribution can be considered the same, even for very large numbers of topological charge $l$. However, when the helical-phase is distinguishable (solid lines), only the rotational velocity for $l = 1$ lays in the same value as the $\omega_{indistinguishable}$. For $l > 1$, there is a notable difference between both rotation speeds.

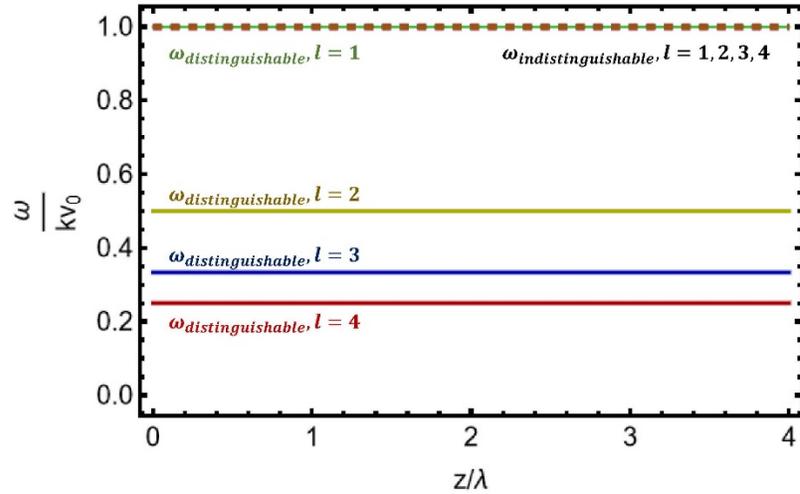

**FIGURE 3.** Rotational velocities $\omega_{indistinguishable}/kv_0$ (dashed lines) and $\omega_{distinguishable}/kv_0$ (solid lines) for $l = 1,2,3,4$ as function of $z/\lambda$, for 633 nm wavelength, beam waist $\omega_0 = 5.0$ mm, and fixed $p = 0$.

## CONCLUSION

A novel approach to identify the distinguishability characteristic of a helical-phase distribution of LG vortex is presented. This feature enables to identify each helical-phase structure in the LG beam phase distribution, presenting relevant impact on the rotational speed of this phase. The simulation of the rotation of the helical phase is performed. The results of such simulation validate the mathematical expressions of the proposed approach. This rotation can be clockwise when the longitudinal phase is taken as exp ($ikz$), and counterclockwise when exp ($-ikz$). If the helical-phase distribution, in vortex beam, cannot be distinguished, each phase distribution would be interpreted as a unique distribution, due to its rotational symmetry, as shown in Fig. 2a.

On the other hand, if the helical structure is distinguished, the angular rotation speed is found using Eq. (6). The larger the azimuthal index, the lower the rotation speed at which the phase distribution rotates. One future applications of the present helical-phase distinguishable theory are the possibility of being used as a coding communication technology to carry some kind of information in each helix from the helical structure of the LG beam phase. To accurately decode the information, it is important to distinguish each helix in the received phase distribution, i.e., we need to know which helix is the first one, the second one, and so on. Therefore, a track of the phase distribution needs to be done while the LG beam moves from the emitter to the receiver. Additionally, using this coding technology in combination with other optical dimensions, can improve more the coding efficiency and increase the information transmission capacity [42]. Another application can be in the optical particle manipulation field, by studying the dynamics of particles trapped in an off beam, including acceleration and deacceleration of the angular rotation speed [43].